\input epsf.tex    

\documentstyle[11pt]{article}

\newcommand{\bmat}{\left(\begin{array}}
\newcommand{\emat}{\end{array}\right)}

\def\yzero{\smash{\hbox{$y\kern-4pt\raise1pt\hbox{${}^\circ$}$}}}

\def\beq{\begin{equation}}
\def\eeq{\end{equation}}
\def\beqa{\begin{eqnarray}}
\def\eeqa{\end{eqnarray}}

\def\-{\hphantom{-}}
\def\ov{\overline}

\def\s2{\frac{1}{2}}

\def\beq{\begin{equation}}
\def\eeq{\end{equation}}
\def\beqa{\begin{eqnarray}}
\def\eeqa{\end{eqnarray}}

\def\IF{\relax{\rm I\kern-.18em F}}
\def\II{\relax{\rm I\kern-.18em I}}

\def\cp{{\cal P}}
\def\IC{\bf C}
\def\IZ{\bf Z}
\def\IR{\bf R}

\def\IS{\bf S}
\def\IP{\bf P}
\def\IT{\bf T}

\def\z2z2{$\IC^3/(\IZ_2\times\IZ_2)$}

\def\id{{\bf 1}}

\def\Dfive{$\widehat{D5}$}
\def\Dseven{$\widehat{D7}$}
\def\s12{\frac 12}

\def\Dsl{\,\raise.15ex\hbox{/}\mkern-13.5mu D} 

 \def\cp#1{\relax\ifmmode {\IP\kern-2pt{}_{#1}}\else $\IP\kern-2pt{}_{#1}$\=fi}

\begin{document}

\makeatletter \@addtoreset{equation}{section} \makeatother
\renewcommand{\theequation}{\thesection.\arabic{equation}}

\pagestyle{empty}
\vspace*{.5in}
\rightline{CERN-PH-TH/2005-190}
\rightline{IFT-UAM/CSIC-05-41}
\rightline{\tt hep-th/0510073}
\vspace{1.5cm}

\begin{center}
\LARGE{\bf From F/M-theory to K-theory and back} \\[10mm]

\medskip

\large{I\~naki Garc\'{\i}a-Etxebarria$^\dagger$, Angel M. 
Uranga$^\ddagger$} 
\\
{\normalsize {\em $^\dagger$ Instituto de F\'{\i}sica Te\'orica, C-XVI \\
Universidad Aut\'onoma de Madrid \\
Cantoblanco, 28049 Madrid, Spain \\
$^\ddagger$ TH Unit, CERN, \\
CH-1211 Geneve 23, Switzerland \\
{\tt innaki.garcia@uam.es, angel.uranga@cern,ch} 
\\[2mm]}}

\end{center}

\smallskip

\begin{center}
\begin{minipage}[h]{14.5cm}
{\small
We consider discrete K-theory tadpole cancellation conditions in type 
IIB orientifolds with magnetised 7-branes. Cancellation of K-theory charge 
constrains the choices of world-volume magnetic fluxes on the latter.
We describe the F-/M-theory lift of these configurations, where 7-branes 
are encoded in the geometry of an elliptic fibration, and their magnetic 
quanta correspond to supergravity 4-form field strength fluxes. In a K3 
compactification example, we show that standard quantization of 4-form 
fluxes as integer cohomology classes in K3 automatically implies the 
K-theory charge cancellation constraints on the 7-brane worldvolume  
magnetic fluxes in string theory (as well as new previously unnoticed 
discrete constraints, which we also interpret). Finally, we show that 
flux quantization in F-/M-theory implies that 7-brane world-volume flux 
quantization conditions are modified in the presence of 3-form fluxes.
}
\end{minipage}
\end{center}

\newpage                                                        

\setcounter{page}{1} \pagestyle{plain}
\renewcommand{\thefootnote}{\arabic{footnote}}
\setcounter{footnote}{0}

\section{Introduction}

One of the important insights into the properties of D-branes is that 
their charges are classified by K-theory \cite{minmoore,wittenkth}. 
K-theory charges differ from cohomology charges in discrete 
pieces, which typically correspond to charges of stable non-BPS D-branes 
\cite{nonbps}.

An interesting consequence of the K-theory classification of D-brane 
charge for string compactification, is that the consistency conditions 
should involve cancellation of RR charge in K-theory, rather than in 
cohomology \cite{urakth}. This leads to additional discrete 
constraints, which are related to cancellation of global anomalies in the 
effective lower-dimensional theory, or on suitable D-brane probes 
\cite{urakth}. These discrete constraints lead to non-trivial 
conditions in diverse compactifications with D-branes (see e.g. 
\cite{marchesano,cascur}) and even play an interesting role in building 
phenomenologically realistic models in string theory \cite{ms,sch}.

It is convenient to understand K-theory charge cancellation conditions 
from different viewpoints. This is important, since they are not 
detectable by standard techniques, e.g. the computation of RR 
tadpoles via factorization of amplitudes reproduces only $\IZ$-valued 
charges. Moreover, their proper geometric interpretation seems to require 
the interpretation of RR fields in K-theory \cite{moorewitten}. As 
mentioned above, a possibility is to relate them to cancellation of global 
anomalies on the world-volume of suitable probes \cite{urakth}.

In this paper we address these consistency conditions from a different 
viewpoint. We consider a type IIB orientifold compactification with 
O7-planes and magnetised D7-branes, namely D7-branes carrying non-trivial 
abelian world-volume gauge bundles. The model has non-trivial K-theory 
$\IZ_2$ charge cancellation conditions, which constrain the magnetic flux 
quanta for the D7-brane gauge fields. We reinterpret and derive these 
conditions by lifting the configuration to F-/M-theory on an 
elliptically fibered K3, where the magnetic fields correspond to 
non-trivial 4-form fluxes supported on 
special loci (the degenerations of the elliptic fibration) in a purely 
geometrical background without branes. We show that proper quantization of 
the 4-form flux, as integer cohomology classes in K3, automatically leads 
to type IIB magnetic fluxes obeying the standard quantization conditions 
plus the additional K-theory $\IZ_2$ constraint. In addition, we find 
new $\IZ_2$ constraints which we properly interpret.

Moreover, we generalize the result in the presence of bulk 4-form fluxes, 
leading to a modification of the $\IZ_2$ quantization conditions on 
D7-brane magnetic fluxes in the presence of non-trivial 3-form fluxes
in the type IIB orientifold. Hence, our results have important 
consequences in the construction of consistent compactifications of type 
IIB theory with 3-form fluxes and D-branes (see \cite{drs,gkp} for a 
general discussion of flux compactifications of this kind, and 
\cite{blt,cascur} for the formalism to incorporate magnetised D-branes).

This paper is organized as follows. In Section \ref{kthcons} we describe 
the K-theory charge cancellation constraints in certain orientifold models 
with D-branes carrying world-volume magnetic fluxes. We also provide the 
lift of an orientifold with O7-planes and magnetised D7-branes in terms of 
F-/M-theory on $\IT^4/\IZ_2\times \IT^2$ with non-trivial 4-form fluxes 
along certain localized 2-forms. In Section \ref{k3geometry} we describe 
the homology of K3 at the $\IT^4/\IZ_2$ point, and construct a generating 
set of 2-cycles in terms of the structure of $\IT^4/\IZ_2$ as an elliptic 
fibration. In Section \ref{quantiz} we use this information to show that 
4-form flux quantization in cohomology in F-/M-theory implies the standard 
quantization of IIB D7-brane world-volume magnetic fluxes, plus additional 
discrete $\IZ_2$ conditions, which correspond to K-theory charge 
cancellation conditions. In Section \ref{threeformflux} we show the 
modification of these conditions when type IIB bulk 3-form fluxes are 
introduced. Finally, Section \ref{final} contains our conclusions.

\section{K-theory constraints in magnetised D-branes}
\label{kthcons}

\subsection{K-theory constraints on type I}

Consider type I string theory compactified on $\IT^2$, with a 
\Dseven-brane 
spanning the eight non-compact dimensions and sitting at a point on 
$\IT^2$. As discussed in \cite{urakth}, this configuration is inconsistent 
due to non-cancellation of the K-theory $\IZ_2$ charge carried by the 
\Dseven-brane. The simplest way to show the inconsistency is to introduce 
a D5-brane probe wrapping the $\IT^2$. The field theory on the 4d 
non-compact dimensions of the probe has a gauge group $USp(2)=SU(2)$. Due 
to the presence of the \Dseven-brane in the background, there is one Weyl 
fermion in the fundamental representation from the $5\hat{7}$ open string 
sector. Hence the probe world-volume 
theory is anomalous due to a Witten global gauge anomaly 
\cite{wittenglobal}. This reflects an inconsistency in the background. 
Compactifications with an even number of \Dseven-branes have however 
proper cancellation of $\IZ_2$ charges and are hence consistent.

This line of argument can also show the existence of $\IZ_2$ constraints 
in type I compactifications without explicit \Dseven-branes, but with 
non-trivial world-volume gauge bundles on the D9-branes. The simplest 
situation is to consider a $\IT^2$ compactification of type I, with 
abelian monopole backgrounds. Namely, denote $U(1)_a$ the 16 Cartan
generators of $SO(32)$, or equivalently, from the viewpoint of the parent 
IIB theory, the $U(1)$ gauge factors carried by the $a^{th}$ D9-brane and 
its orientifold image (denoted $a'$). Consider introducing a constant 
$U(1)_a$ magnetic field $F_a$ on $\IT^2$, 
satisfying the quantization condition 
\beqa
\int_{\IT^2} F_a = n_a
\label{quant1}
\eeqa
These are the so-called magnetised D-brane constructions, studied among 
others in \cite{bachas, bgkl, aads,afiru}. Notice that there is an additional 
freedom in choosing the multiwrapping of the D-branes on the $\IT^2$ 
\cite{bgkl}, but for simplicity we will not include it, since it is not 
essential to our main point.

The magnetic field induces $n_a$ units of IIB D7-brane charge on the 
$a^{th}$ D9-brane, and of $\ov{D7}$-brane charge on the $a'^{th}$ image 
D9-brane. Given that a D7-$\ov{D7}$ pair descends to a \Dseven-brane 
of the quotient type I theory \cite{wittenkth}, the compactification 
contains 
an overall number of $\sum_a n_a$ units of \Dseven-brane charge. 
Cancellation of the K-theory $\IZ_2$ charge thus requires
\beqa
\sum_a n_a \in 2\IZ
\label{cancel1}
\eeqa
Again, the inconsistency of the backgrounds which do not satisfy the above 
constraint can be made manifest by introducing a D5-brane probe wrapped on 
$\IT^2$. Namely, in the 59 open string sector the index of the Dirac 
operator coupled to the gauge bundle is odd, leading to an odd number of 
4d Weyl fermions in the fundamental representation of the symplectic group 
on the D5-branes.

The generalization of the above discussion is straightforward. In any type 
I compactification on a manifold $X$, with D9-branes (and images) carrying 
an $SO(32)$ gauge bundle $V$, the K-theory charge corresponding to induce 
\Dseven-branes should cancel. Regarding the type I real bundle in terms 
of type IIB D9-branes carrying a $U(16)$ gauge bundle $E$, and their 
images, cancellation of the K-theory induced \Dseven-brane $\IZ_2$ charge 
requires $c_1(E)$ to be an even class in $H^2(X,\IZ)$.

\subsection{K-theory constraints on type IIB orientifolds}
\label{kthoniib}

Let us return to the type I $\IT^2$ compactification, and obtain similar 
conclusions for related orientifold models. Consider compactifying on a 
further $\IT^2$ (denoted $(\IT^2)'$ to avoid confusion), and T-dualizing 
on its two directions. We obtain type IIB on $(\IT^2)'\times \IT^2$, 
modded out by $\Omega R(-1)^{F_L}$, where $R$ flips the two coordinates 
of $(\IT^2)'$ and leaves $\IT^2$ invariant. There are four O7-planes 
sitting at the fixed points in $(\IT^2)'$, and 16 D7-branes and their 
images, with world-volume magnetic fields $F_a$ satisfying 
the quantization (\ref{quant1}).

The above arguments show that the configuration must satisfy the 
additional $\IZ_2$ cancellation condition (\ref{cancel1}). By T-duality, 
the $\IZ_2$ charge is now associated to a non-BPS D5-brane (\Dfive-brane) 
wrapped on $(\IT^2)'$. This note is devoted to developing a better 
understanding of the additional $\IZ_2$ condition in this system.

What is interesting about this configuration is that it has a direct 
connection with other dual descriptions in terms of compactifications of 
F/M-theory. Moreover, it contains several of the 
ingredients present in many of the recently studied flux compactifications 
with O3-planes, D3-branes, and 7-branes \cite{drs,gkp}. In fact, the 
relation will be far more direct once we study the lift to F/M-theory, in 
the next section.

\subsection{Lift to F/M-theory with fluxes}

In this section we describe the lift of the above configurations to 
F/M-theory, and investigate the interpretation of the $\IZ_2$ K-theory 
constraint in the latter context.

The lift of type IIB orientifolds with O7-planes and D7-branes, in the 
absence of world-volume magnetic fields is familiar and has been discussed 
e.g. in \cite{sen}. One obtains F/M-theory 
\footnote{The relation to M-theory arises upon an additional 
compactification on a circle. Since the background (geometry plus 
fluxes) is most physical in the M-theory setup, we center on this picture 
when describing it. Note however that the M-theory backgrounds 
surviving in the F-theory limit are not the most general ones. We use
the notation F/M-theory to refer to M-theory on backgrounds of the former 
kind.}
compactified on an elliptic 
fibration over a base given by the IIB compactification space. 
Degenerations of the elliptic fibration where a $(p,q)$-cycle degenerates 
correspond to $(p,q)$ 7-branes in the IIB picture. As discussed in 
\cite{sen}, the lift of O7-planes correspond to a set of two fiber 
degenerations, whose $(p,q)$ label can be chosen \cite{gz} to be $(1,-1)$ 
and $(3,-1)$, in the convention where a D7-brane lifts to a $(1,0)$ 
degeneration.

The world-volume gauge field on the 7-branes arises as follows. The local 
geometry around the pinching point of a degenerate elliptic fiber is that 
of a Taub-NUT geometry. This geometry supports a normalizable harmonic 
2-form \cite{senkk}, which we denote by $\omega$. In the M-theory picture, 
the 
Kaluza-Klein reduction of the 3-form leads to a component of the type
\beqa
C_3 \, = \, \omega \wedge A_1
\eeqa
where $A_1$ is a 1-form supported on the degeneration locus on the base, 
namely on the 7-brane world-volume. Hence, it corresponds to the gauge 
field carried by the 7-brane. Therefore, the world-volume magnetic fluxes
on a 7-brane correspond to components of the 4-form field strength, of the 
type
\beqa
G_4 \, = \, \omega \wedge F_2
\eeqa
Namely, they are mapped to standard supergravity 4-form fluxes in an 
otherwise purely geometrical compactification. Hence, interestingly the 
question of the constraints (quantization and additional $\IZ_2$ 
condition) on the IIB 7-brane magnetic fluxes is intimately related to 
the quantization conditions of 4-form fluxes in F/M-theory 
compactifications. 

The setup corresponds to compactifications of F/M-theory with 
non-trivial $G_4$ field strength fluxes, which have been under intense 
study by themselves on in their type IIB avatar \cite{beckers,drs,gkp}. 
For the moment we simply consider $G_4$ fluxes which receive the 
interpretation of IIB gauge backgrounds, leaving the discussion of $G_4$ 
fluxes which reduce to IIB 3-form fluxes for Section 
\ref{threeformflux}.

One may wonder if quantization of $G_4$ as an integer cohomology class is 
sufficient to guarantee the consistency of the backgrounds, or if some 
additional discrete constraint is required in order to account for the 
extra $\IZ_2$ condition in string theory. In coming sections we study
a particular compactification on K3, which is simple enough for detailed 
analysis, but still keeps all the required subtle structure. We show that 
standard integer quantization of $G_4$ automatically implies the appropriate 
constraints on the IIB 7-brane world-volume magnetic fluxes, including 
the above mentioned $\IZ_2$ constraint.

\subsection{Gauge fields for K3 at the $SO(8)^4$ point}

In an F-theory compactification, components of the 3-form along 2-forms 
localized at pinchings of the elliptic fiber correspond to gauge fields on 
the corresponding IIB 7-branes. However, not all such localized 2-forms 
are independent, and the total number of gauge fields is smaller than the 
number of degenerate fibers \cite{vafafth}. Our strategy in the present 
paper is to pick a set of basic degenerations, 
which generate all gauge fields in the theory, so that one can safely 
ignore the 2-forms supported on the remaining degenerations (which can be 
regarded as non-dynamical). 

This splitting is manifestly natural and justified in 
F/M-theory models which correspond to lifts of type IIB orientifold with 
local cancellation of 7-brane charge, namely with sets of exactly four 
D7-branes on top of each O7-plane. In the corresponding lift, we 
locally find a Kodaira $I_0^*$ degenerate fiber \cite{sen}, to be 
discussed later on. 
Slightly separating the D7-branes from the orientifold plane leads to 
a lift with four $(1,0)$ fibers, corresponding to the D7-branes, and a set 
of $(1,-1)$ and $(3,-1)$ fibers, corresponding to the O7-plane. In this 
setup is it natural to associate the independent 2-forms $\omega_i$ of the 
$(1,0)$ degenerations to the D7-branes, while the 2-forms of the remaining 
two fibers are not independent, and do not generate new gauge fields (as 
corresponds to their interpretation as an O7-plane). Far away from such 
situation, the relation between the gauge fields in the lower-dimensional 
theory and the localized 2-forms is more involved, and we will not enter 
its discussion in the present note.

The simplest situation of the above kind corresponds to the setup in 
Section \ref{kthoniib}, namely type IIB on $(\IT^2)'$ modded out by 
$\Omega R$, with 
8 D7-branes (four and their four images) on top of each of the four 
O7-planes. The gauge symmetry from the D7-branes corresponds to $SO(8)^4$.
The F-theory lift corresponds to a compactification on the $\IT^4/\IZ_2$, 
orbifold limit of K3. The appearance of the gauge symmetry in this setup 
is described below. We refer to this locus in the moduli space of the 
system as the $SO(8)^4$ point.

Following the above general strategy, the F/M-theory lift of the 
Cartan gauge fields on the 16 D7-branes at the $SO(8)^4$ point, and its 
corresponding field strengths, correspond to supergravity backgrounds for 
the M-theory 3-form $C_3$ and its field strength $G_4$
\beqa
C_3\, =\, \sum_{a=1}^{16}\, \omega_a \wedge A_1^{(a)} \quad , \quad
G_4\, =\, \sum_{a=1}^{16}\, \omega_a \wedge F_2^{(a)} 
\eeqa
where $\omega_a$ label the 2-forms supported on the degenerations 
corresponding to D7-branes in the IIB reduction. Clearly, the 2-forms 
associated to other degenerate fibers do not lead to dynamical fields in 
the IIB reduction, hence showing they are not independent degrees of 
freedom.

\section{Geometry of K3 at the $SO(8)^4$ point}
\label{k3geometry}

\subsection{The homology lattice of $\IT^4/\IZ_2$}

A general discussion of the homology lattice of K3 may be found in 
\cite{aspk3}. Here, as mentioned, we consider for concreteness the K3 at 
the $SO(8)^4$ point,
to make the connection with the type IIB orientifolds more manifest \cite{sen}. 
Hence we consider the K3 to be given by the orbifold $\IT^4/\IZ_2$, which 
for simplicity we also take to be rectangular. 

A basis of  $H_2(\IT^4/\IZ_2,\IR)$, which is 22-dimensional, is provided 
by the following set of 2-cycles: We first have the six independent 
2-cycles on $\IT^4$, which are invariant under the $\IZ_2$ action. 
We also have the 16 two-spheres arising from blowing up the 
16 local $\IC^2/\IZ_2$ orbifold singularities.

It is convenient to introduce some notation for these 2-cycles. Let us 
label the four coordinates in $\IT^4$ by latin indices $i=1,2,3,4$. We 
denote $\Pi_{ij}$ (antisymmetric in the two indices) the toroidal 2-cycle 
spanning the directions $i,j$. The 
sixteen blowup 2-cycles sit at points with coordinates $0$ or $1/2$ 
along each of the four coordinates (in units of the corresponding radius). 
Denoting these two possible values by $+$ and $-$, we denote the blowup 
2-cycles by $e_{a_1 a_2 a_3 a_4}$, where $a_i=+,-$ for $i=1,2,3,4$ denotes 
the location of the $\IZ_2$ fixed point in the $i^{th}$ direction in 
$\IT^4$.

It is easy to realize that the above set of 2-cycles, which span the 
so-called Kummer lattice, do not form the complete lattice of the K3 
second homology. Rather, they form a sublattice of index two (see e.g. 
\cite{aspk3,bbkl} for discussion). Indeed, there exist 2-cycles in the 
quotient, spanning two directions of the $\IT^4$ and passing through 
four $\IZ_2$ fixed points, which have half the volume of the corresponding 
toroidal cycles. Since these 2-cycles are mapped to themselves by the 
$\IZ_2$ action, one may think about them as fractional 2-cycles 
\footnote{This is convenient in that a brane wrapped on such 2-cycle is 
described as a fractional brane in orbifold language. In the geometric 
sense it is somewhat of a misnomer, since it belongs to the K3 {\em 
integer} homology lattice.}. It is convenient to introduce some notation 
for these 2-cycles, as follows. Consider e.g. the 2-cycle spanning the 
directions $3,4$, and passing through the fixed points 
associated to $e_{++++}$, $e_{+++-}$, $e_{++-+}$ and $e_{++--}$. We denote 
its class by $f_{++34}$. Clearly $f_{++34}$ and $\Pi_{34}$ are not 
independent, but satisfy the relation 
\beqa
2\, f_{++34}\, +\, e_{++++}\, +\, e_{+++-}\, +\, 
e_{++-+}\, +\, e_{++--}\, =\, \Pi_{34}
\label{rel1}
\eeqa
Clearly, there exist 2-cycles $f_{+-34}$, $f_{-+34}$ and $f_{--34}$ 
spanning the directions $3,4$ but passing through a different set of fixed 
points, and having a similar relation to $\Pi_{34}$. Also, there are 
classes $f$ associated to any other pair of directions. A more explicit 
realization of the 2-cycles $f$ will be manifest in coming paragraphs.

The intersection numbers of these cycles can be obtained from
\beqa
\Pi_{ij}\cdot \Pi_{kl} & = & \epsilon_{ijkl} \nonumber \\
\Pi_{ij}\cdot e_{a_1a_2a_3a_4} & = & 0 \nonumber \\
e_{a_1a_2a_3a_4}\cdot e_{b_1b_2b_3b_4} & = & -2\,
\delta_{a_1b_1}\delta_{a_2b_2}\delta_{a_3b_3}\delta_{a_4b_4}
\eeqa
and relations similar to (\ref{rel1}). For instance we have 
$f_{++34}\cdot e_{++a_3a_4}=1$, etc.

\subsection{Homology of $\IT^4/\IZ_2$ as en elliptic fibration}

\begin{figure}
\begin{center}
\epsfxsize=10cm
\hspace*{0in}\vspace*{.2in}
\epsffile{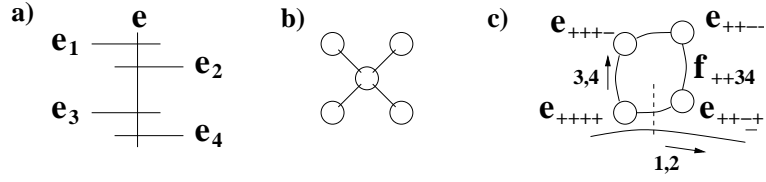}
\caption{\small a) The non-trivial 2-cycles in a Kodaira $I_0^*$ singular 
fiber b) the Dynkin diagram of $SO(8)$ c) realization over a degeneration 
point in $\IT^4/\IZ_2$.}
\label{so8}
\end{center}  
\end{figure}

In order to connect with the type IIB orientifold model, it is 
convenient to describe the homology lattice of this K3 in terms of its 
interpretation as an elliptic fibration. Consider the torus 
spanned by the coordinates 34 (namely $\Pi_{34}$) to be the elliptic 
fiber. The fibration has four degenerate fibers, located over the points 
of the base $\IT^2$ fixed by the $\IZ_2$ action. As described in 
\cite{sen}, the degenerate fibers are of Kodaira type $I_0^*$, and lead, 
in the F-theory limit of shrinking fiber class, to an enhanced $SO(8)$ 
gauge symmetry. Indeed (see e.g. \cite{aspk3}) the (extended) Dynkin 
diagram of this symmetry is realized in terms of the 2-cycles of the 
reducible $I_0^*$ fiber, as shown in figures \ref{so8}a,b. This diagram 
encodes the intersection numbers among the classes, namely 
\beqa
e\cdot e=-2 \quad , \quad e_\alpha\cdot 
e_\beta=-2\delta_{\alpha\beta}\quad, \quad e\cdot 
e_\alpha=1
\label{inumbers}
\eeqa
The 2-cycles are 
related to the fiber class by 
\beqa
2e+e_1+e_2+e_3+e_4=f
\eeqa

The relation of these 2-cycles with the above general description is 
clear. Let us center on a particular degeneration point on the base, e.g. 
that corresponding to the origin in the coordinates $1,2$. Using our 
general description above, on top of this point we have the class 
$f_{++34}$, and the blowup classes $e_{++ab}$. Their intersection numbers, 
along with the relation (\ref{rel1}), leads to the relation
\beqa
f_{++34}=e \quad , \quad e_1=e_{++++}\quad , \quad e_2=e_{+++-} \quad, 
\quad e_3=e_{++-+} \quad ,\quad e_4= e_{++--}
\eeqa
as illustrated in figure \ref{so8}c. The above relation is unique up to 
the action of the Weyl group, and we will use the above convention in the 
remainder of the paper. Namely, $\alpha$ is a double index $(a_3,a_4)$ 
denoting the position on the fiber, with $\alpha=1,2,3,4$ corresponding to 
$a_3a_4=++,+-,-+,--$.

The above discussion shows that the class $e$ is 
precisely one of the classes required to refine the Kummer lattice to the 
actual integer homology lattice of K3. Similar discussions apply to the 
classes $f_{a_1a_234}$.

\medskip

It is also possible to provide a description of the remaining classes $f$
in terms of the elliptic fibration picture. The simplest are the classes 
$f_{12a_3a_4}$, which correspond to different sections of the torus 
fibration \footnote{Since an elliptic curve is a torus with a choice of a 
point, the elliptic fibration implies the choice of one of these sections 
as the base.}. The remaining classes e.g. $f_{1a_23a_4}$ are related to 
the 2-torus classes $\Pi_{13}$, etc. In the elliptic fibration, 2-cycles like 
$\Pi_{13}$ are realized by considering a circle on the base, spanning the 
direction $1$, and fibering over it the circle of the fiber corresponding 
to $3$. An important point is that the circle on the base surrounds two 
$I_0^*$ degenerations, each of them leading to an $SL(2,\IZ)$ monodromy 
$-\id_2$, so the total monodromy along the base circle is trivial, and the 
circle along $3$ can be consistently fibered to form a 2-cycle. Actually, 
this representative of $\Pi_{13}$ can be moved toward the two $I_0^*$ 
degeneration points (corresponding e.g. to the coordinate positions $(++)$ 
and $(-+)$ on the base), where it becomes reducible via
\beqa
\Pi_{13} \, = \, 2f_{1+3+}\, + \, e_{++++}\, +\, e_{++-+}\, +\, e_{-+++}
\, +\, e_{-+-+}
\eeqa
Namely, the class $f_{1+3+}$ corresponds to a 2-cycle spanning the 
directions $1,3$, located at origin in the direction $2,4$, and passing 
through the above four fixed points.
Clearly one can operate similarly to describe the remaining classes of 
type $f$.

\section{Type IIB gauge field flux quantization from F/M-theory}
\label{quantiz}

In this section we employ the quantization conditions on 4-form fluxes in 
F/M-theory to obtain the corresponding quantization conditions for the IIB 
gauge field fluxes. The key point in the analysis is the computation of 
the periods of the 2-forms $\omega$ associated to the D7-branes. The 
bottom line is that these 2-forms are not integer cohomology classes, but 
suitable linear combinations of them are. As we show this implies that 
appropriate quantization of $G_4$ imposes the extra constraint on the 
D7-brane gauge field fluxes (\ref{cancel1}), as well as some previously 
unnoticed ones. 

\subsection{The forms $\omega$ revisited}

In this section we determine the integrals of the forms $\omega$, involved 
in the definition of the D7-brane gauge fields, over different 2-cycles in 
the homology lattice of K3. A simple way to obtain these integrals is to 
compute the charges of BPS states (corresponding to branes wrapped on such 
2-cycles) under the Cartan generators of the associated $SO(8)$ gauge 
symmetries. 

{\bf Cycles localized at $I_0^*$ fibers}

The realization of the homology classes of $\IT^4/\IZ_2$ from the elliptic 
fibration viewpoint makes the relation to the type IIB orientifold 
manifest. In particular, it is possible to realize the 2-cycles 
associated to the above classes $e$, $e_i$ in an $I_0^*$ fiber in terms 
of string junctions \cite{junctions} in the IIB orientifold. Hence 
junction techniques allow to compute the integrals of the 2-forms 
$\omega_i$ over those 2-cycles. Physically this amounts to computing
the charges of the BPS states corresponding to M-theory M2-branes wrapped 
on those 2-cycles. In junction language, they are simply given by the 
numbers of string prongs ending on the relevant D7-branes, counted with 
orientation.

\begin{figure}
\begin{center}
\epsfxsize=10cm
\hspace*{0in}\vspace*{.2in}
\epsffile{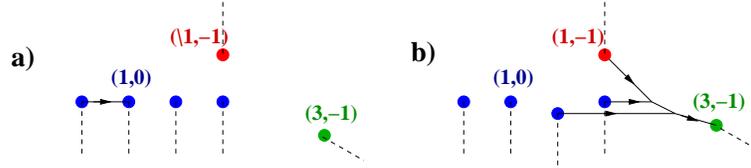}
\caption{\small Some junctions representing BPS states near a Kodaira 
$I_0^*$ singularity. Labels denote the $(p,q)$ type of degeneration of 
points of the corresponding color, with the blue $(1,0)$ nodes 
corresponding to D7-branes in the IIB interpretation. Figure a) 
represents the state corresponding to the $SO(8)$ non-zero roots of the 
form $(+1,-1,0,0)$. Figure b) represents the state corresponding to the 
non-zero root $(0,0,+1,+1)$. For clarity, one of the D7-branes has been 
slightly shifted downward. The remaining non-zero roots are obtained by 
choosing similar junctions involving different pairs of D7-branes.}
\label{bps}
\end{center}  
\end{figure}

The junctions for the 2-cycles of type $e$, $e_i$, are shown in figure 
\ref{bps}, and it can be checked that their intersection numbers 
(determined using junction rules) agree with (\ref{inumbers}). The charges 
of the corresponding BPS states with respect to the four $SO(8)$ Cartan 
generators, or equivalently the integrals of the different 2-forms 
$\omega_i$ over a given 2-cycle, can be described by a vector
\beqa
e_1=(1,-1,0,0)\; ,\; 
e_2=(1,1,0,0)\; ,\; 
e_3=(0,0,1,-1)\; ,\; 
e_4=(0,0,1,1)\; ,\; 
e=(-1,0,-1,0) \quad   
\label{i0weights}
\eeqa
The $SO(8)$-root structure of these vectors is manifest.

\medskip

{\bf Additional 2-cycles}

In the previous discussion we have determined the integrals of the forms 
$\omega_i$ associated to a given $SO(8)$ along the 2-cycles localized on 
that degenerate fiber. However, it is important to know the integrals over 
other 2-cycles in the K3. As usual, this can be done by wrapping M2-branes 
on the corresponding M-theory 2-cycles and computing the charges of the 
corresponding states under the Cartan generators of the $SO(8)$. 

Let us consider a IIB D1-brane passing through two stacks of O7/D7-branes, 
e.g. with coordinates $++$ and $-+$ in the directions $1,2$ 
(being fixed under the orientifold action, this is a half D1-brane). 
The charges of this state under the 
Cartan generators of the two $SO(8)$'s through which it passes are easily 
computed in the IIB picture. In the sector of 1-7 open strings, there are 
fermion zero modes in the vector representation of the $SO(8)$'s. Upon 
quantization of these zero modes, the object transforms as a spinor 
simultaneously under both $SO(8)$'s. The overall chirality is fixed by the 
GSO projection, so the state transforms in the $(8_s,8_s)+(8_c,8_c)$. 
Namely, the weight vector can be written
\beqa
(\pm \s12, \pm \s12, \pm \s12, \pm \s12; \pm \s12, \pm \s12, \pm \s12,\pm \s12)
\label{spinorweight}
\eeqa
with a total even number of minus signs. This gives the charge of the 
(multiplet of) states under the Cartan generators of the two $SO(8)$'s. 
Notice that different states in a multiplet are related by addition of 
vectors of type (\ref{i0weights}).

The above set of charges can be understood from the geometric viewpoint as 
follows. From the F/M-theory viewpoint the above D1-brane state 
corresponds to an M2-brane wrapped on a cycle spanning the directions 1, 
3, for instance $f_{1+3+}$. From the intersection numbers with the 
exceptional classes $e_{++a_3a_4}$ and $e_{-+a_3a_4}$, and the expression 
of the latter in terms of vectors (\ref{i0weights}), it follows that an
M2-brane on $f_{1+3+}$ has a charge vector
\beqa
(-\s12,\s12,-\s12,\s12;-\s12,\s12,-\s12,\s12)
\eeqa
under the Cartan generators of the $SO(8)$ factors located at 
$(a_1,a_2)=(+,+), (-,+)$.
By considering M2-brane on other holomorphic 2-cycles, differing from 
$f_{1+3+}$ or $f_{1+3-}$ by the addition of exceptional classes, one can 
generate the whole multiplet
\footnote{Actually, in order to generate the whole multiplet, we need to 
consider the addition of e.g. the class $f_{1++4}$ mentioned below. 
This simply means considering D1-branes with some induced F1 charge.}
of states (\ref{spinorweight}).

This implies that the integrals of the forms $\omega_i$ of the 
corresponding $I_0^*$ degenerations are half-integers. Notice again that 
this does not imply any contradiction with the integer intersection numbers 
between 2-cycles in the integer lattice of K3. Clearly, similar states 
can be constructed for any pair of $SO(8)$ factors, leading to bi-spinors 
of the corresponding $SO(8)^2$ groups.

Finally, we can also consider fundamental strings stretching through pairs 
of $SO(8)$ groups. This leads to states in the representation 
$(8_v,8_v)$ under them. In the M-theory realization, we have M2-branes 
wrapped on e.g. $f_{1++4}$. The intersection numbers of this class with 
$e_{++a_3a_4}$ and $e_{-+a_3a_4}$ lead to a set of charges 
\beqa
(-1,0,0,0;-1,0,0,0)
\eeqa
Again, the rest of the multiplet may be obtained by adding exceptional 
classes. Hence the integrals of the forms $\omega_i$ over those cycles are 
integers.

\subsection{Quantization conditions}

The above characterization of the forms $\omega_i$ is sufficient to
understand the relation between flux quantization in 
F/M-theory and string theory, as we do in this section.

Start with a set of D7-brane magnetic fields $F_{A,i}$, where $A$ labels 
the $SO(8)$ factor, and can be regarded as a double index $A=(a_1,a_2)$ 
encoding the coordinates in the directions $1,2$. This configuration can 
be lifted to F/M-theory as
\beqa
G_4=\sum_A \sum_i \omega_{A,i} \wedge F_{A,i}
\eeqa
The requirement that the flux $G_4$ is quantized over 4-cycles of the kind 
$e_A\times \IT^2$, where $e_A$ is an exceptional class over the $A^{th}$ 
$I_0^*$ fiber, implies that
\beqa
\int_{\IT^2} F_{A,i} = n_{A,i} \in \IZ
\eeqa
which is the familiar quantization of D-brane world-volume magnetic 
fields. In addition, we have the conditions that the flux $G_4$ is 
quantized over 4-cycles of the kind $f \times \IT^2$, where $f$ is a 
2-cycle passing through the elliptic degeneration $A$ and $B$ on the 
base. For instance, $f_{1+3+}$ for $A=(++)$, $B=(-+)$. These conditions
imply
\beqa
\sum_i\, (\, n_{A,i}\, +\, n_{B,i}\, ) \, \in \, 2 \IZ
\label{stronger}
\eeqa
This should hold for any pair $A,B$. Clearly by combining these conditions 
one can obtain the constraint
\beqa
\sum_{i,A}\, n_{A,i}  \in 2\IZ
\label{weaker}
\eeqa
where the sum clearly runs over all $U(1)$ magnetic fields in the theory. 
This is the condition (\ref{cancel1}), which arises from K-theory in the 
IIB language. In our setup, it has automatically arisen from proper 
quantization of fluxes in F/M-theory. 
In other words, the $\omega$'s are not integer cohomology classes, but 
suitable linear combinations of them are. The requirement that $G_4$ is an 
integer cohomology class forces the magnetic flux quanta to adjust so that 
one recovers linear combinations of $\omega$'s which are integer classes.

This result nicely agrees with the 
observation in \cite{dmw} that K-theory in string theory arises from 
homology in M-theory. 

\subsection{New constraints on gauge field fluxes}
\label{newconstraints}

The conditions (\ref{stronger}) imply (\ref{weaker}), but are actually 
stronger. We conclude this section with a discussion of these new $\IZ_2$
conditions on the quantization gauge field fluxes arising 
the 4-form flux quantization. 

In type IIB string theory, these conditions do not correspond to 
cancellation of a cohomological charge, hence they should correspond to 
cancellation of a K-theory charge. Indeed, let us consider making $\IT^2$ 
non-compact and studying the structure of the world-volume gauge bundle 
over the resulting non-compact 2-plane. In this setup, the quantity
\beqa
\sum_{i=1}^4 \, \int_{\IR^2}\,  F_{A,i} \, +\, \sum_{i=1}^4 \, 
\int_{\IR^2}\, F_{B,i}
\eeqa
mod 2 corresponds to a non-trivial $\IZ_2$ charge carried by the 
bundles. This $\IZ_2$ charge can be measured by considering the particle 
obtained by wrapping a D1-brane on the fixed circle passing through the 
$A^{th}$ and $B^{th}$ fixed points (hence transforming as a spinor under 
both $SO(8)$'s), and moving it around the circle at 
infinity in the 2-plane. The condition (\ref{stronger}) simply amounts to 
requiring cancellation of this K-theory charge in the compact context.

Notice the intricate relation between the existence of the
non-trivial $\IZ_2$ charge and the global structure of the gauge group 
(which precisely allows the existence of the above bi-spinor state). Hence 
the above statement depends crucially on the global structure of the 
gauge group, which can be described as follows. Define the $Spin(8)^4$ 
element $x_{AB}$ as that acting as $-1$ on the representation 
$(8_v,1)+(1,8_v)$ and $(8_s,8_c)$ and $(8_c,8_s)$ of the $A^{th}$ and 
$B^{th}$ $Spin(8)$ factors, and as $+1$ on $(8_s,8_s)$, $(8_c,8_c)$ and 
the adjoint representations. The set of $x_{AB}$'s form a subgroup 
$\Gamma$ 
of $Spin(8)^4$. The gauge group of the our compactification is globally 
$Spin(8)^4/\Gamma$, so that there is an obstruction to define states in 
the representations picking minus signs under these actions \footnote{The 
obstruction to define such states could be defined as the generalized 
Stiefel-Whitney class in \cite{wittenvector} obstructing vector 
structures in the different $Spin(16)$'s containing $Spin(8)_A\times 
Spin(8)_B$.}, and such states are indeed absent from the theory. On the 
other hand, states transforming in the $(8_s,8_s)$ or $(8_c,8_c)$ can be 
(and are) present, and can be used to measure the $\IZ_2$ charge carried 
by the bundles. Hence, the classification of K-theory classes of bundles 
with this global group structure should contain the $\IZ_2$ charges 
encountered above. In a compact setup, consistency requires the 
cancellation of such $\IZ_2$ charges.

Finally, let us mention that the global structure of the gauge group 
(and hence the associated K-theory) makes it clear from the string 
viewpoint that there are no additional discrete constraints. Namely, 
ignoring the global structure of the gauge group one might have been 
tempted to interpret the quantity $\sum_{i} n_{A,i}$ mod 2 as a discrete 
charge carried by each $SO(8)$ factor, which should therefore be canceled 
in a compact setup. However, the only way to detect this quantity as a 
physical charge would be to probe the theory with a state in the spinor of 
just one of the $Spin(8)$ factors. Such states are not present in the 
theory (and in fact are not allowed by the global structure of the group), 
hence the above quantity is not a physical charge of the system, and we 
need not require its cancellation, even in a compact setup.

\section{Adding 3-form fluxes}
\label{threeformflux}

The lift of world-volume magnetic fluxes on D7-branes naturally leads to 
M-theory compactifications with non-trivial $G_4$-flux along the localized 
forms $\omega$. It is therefore natural to discuss the situation with more 
general $G_4$-fluxes. As discussed in \cite{drs,gkp}, the only additional 
components of $G_4$ which survive in the F-theory limit are those with one 
leg in the elliptic fiber, and correspond to type IIB NSNS and RR 3-form 
fluxes, $H_3$ and $F_3$ respectively. Defining the complex flux 
$G_3=F_3-\tau H_3$, where $\tau$ is the IIB complex coupling (complex 
structure of the fiber torus), we have the relation
\beqa
G_4\, =\, \frac{G_3d\bar{w}}{\bar{\tau}-\tau}\, +\, {\rm h.c.}
\eeqa
where $dw=dx_4+idx_3$. In our situation of square torus, $\tau=i$ and 
\beqa
G_4\, =\, H_3\, dx_4\, +\, F_3\, dx_3
\eeqa
Hence a general configuration of $G_4$-flux corresponds to a general 
configuration of type IIB $G_3$-flux and D7-brane world-volume magnetic 
fluxes.

Given that world-volume and 3-form fluxes in IIB are components of the 
same kind of flux in F/M-theory, it is important to consider the possible 
effect of bulk 3-form fluxes on the quantization conditions for 
world-volume magnetic fluxes. In this section we address this issue and 
find that the conditions are modified.

Clearly, the only components of $G_3$ which may have a non-trivial effect 
of the D7-brane magnetic quanta $n_{a}$ are those with one leg on 
$(\IT^2)'$ and two on $\IT^2$. Namely
\beqa
H_3\, =\, N_{14} dx_1 dx dy \, + \, N_{24} dx_2 dx dy \quad , \quad
F_3\, =\, N_{13} dx_1 dx dy \, + \, N_{23} dx_2 dx dy 
\eeqa
where $dx dy$ is the volume form of $\IT^2$, and the coefficient notation 
and signs are set for convenience. The coefficients $N$ is this expression 
denote the quanta of 3-form flux along toroidal cycles. Since the 
orientifold action actually quotients $(\IT^2)'$ by a $\IZ_2$ action, one 
might think that the flux quanta are forced to be even. However, as we 
show using the F/M-theory picture, the quantization conditions are 
trickier and allow for odd quanta in certain situations. The cases 
including odd quanta are however more subtle and lead to a modification 
of the quantization conditions (\ref{stronger}) for 2-form magnetic field 
fluxes.

In the F/M-theory picture we can parametrize the relevant $G_4$ 
schematically as
\beqa
G_4\, =\, (\, N_{13} dx_1 dx_3 \, +\, N_{23} dx_2 dx_3\, +\,
N_{14} dx_1 dx_4 \, + \, N_{24} dx_2 dx_4 \, +\, \sum_{a=1}^{16} n_a 
\omega_a\, )\wedge dx\, dy
\eeqa
The key point is that forms like $dx_1 dx_3$ are not integer classes 
in K3, since their integrals along 2-cycles like $f_{1+3+}$ measure the 
volume of the latter, and are half-integers, since the volume is half that 
of a toroidal cycle. However, although forms like $dx_1dx_3$ and $w_a$ are 
not integer classes, suitable linear combinations of them are. Hence, 
quantization of $G_4$ implies that the coefficients in the above 
expression must adjust to form such linear combinations.

In the following we center on a particular set of fluxes, which suffices 
to illustrate the main point. Consider 
\beqa
G_4\, =\, (\, N_{13} dx_1 dx_3 \, +\,\sum_{A,i} n_{A,i}  \,
\omega_{A,i} \, )\wedge dx\, dy
\eeqa
Since e.g. $\int_{f_{1+3+}} dx_1dx_3=1/2$, requiring this flux to be 
quantized over the 4-cycle $f_{1+3+}\times \IT^2$ leads to the condition
\beqa
N_{13} \, +\, \sum_{a3,a4} \, (\, n_{++a_3a_4}\, +\, n_{-+a_3a_4}\,) \in 
2 \IZ
\eeqa
Hence for even $N_{13}$ we recover the condition (\ref{stronger}), while 
for odd $N_{13}$ we find a modified quantization condition on the D7-brane 
world-volume magnetic fluxes. Notice that the quantization condition 
(\ref{weaker}) is unchanged.

The general pattern is clear. Given a pair of $SO(8)$ gauge factors, the 
sum of the magnetic flux quanta for their Cartan generators must be even 
(resp. odd) if the number of RR 3-form flux quanta along the 
corresponding 1-cycle (times $\IT^2$) is even (resp. odd) \footnote{This 
kind of condition is similar of that encountered in 
toroidal compactifications with O3-planes, with odd 3-form flux quanta 
\cite{fp,kst}. It would be interesting to understand a possible relation 
between the two mechanisms.}. 

On the other hand, one can use similar computations to show that NSNS 
fluxes do not modify the world-volume magnetic flux quantization 
conditions. This follows from the fact that the integrals of the forms 
$\omega_{A,i}$ over 2-cycles like $f_{1++4}$ are integer.
This also implies that and odd quantum of NSNS flux cannot be 
compensated by a suitable choice of magnetic flux quanta. Therefore, 
proper quantization of the M-theory 4-form flux leads to IIB NSNS fluxes 
quantized to even integers. The asymmetry in the behaviour of NSNS and RR 
fluxes may appear striking, but one should recall that the IIB background 
at hand contains D7-branes, hence is not invariant under S-duality.

The modification of the world-volume flux quantization conditions shows 
that the presence of 3-form fluxes modifies the K-theory of the 
configuration. It would be highly desirable to develop a better 
understanding of this, possible in terms of some twisted K-theory 
\cite{wittenkth}. However, the modification we have found is due to RR 
3-form fluxes, while twisted K-theory has been suggested as the right 
formalism to incorporate NSNS 3-form fluxes. We leave this as an open 
question.

Finally recalling our discussion in Section \ref{newconstraints}, the 
change in the 
quantization conditions reflects a change in the global structure of the 
gauge group. This is the first example of such an effect, as far as we are 
aware of. Again, it would be interesting to gain a better IIB 
interpretation of this effect.

\section{Final comments}
\label{final}

In this paper we have considered the F/M-theory lift of configurations of 
D7-branes with world-volume magnetic fluxes, and described the appearance 
of the quantization conditions on the latter (involving the additional 
subtle K-theory constraints) from quantization of 4-form fluxes as integer 
cohomology forms in the former.

This result fits nicely with the computation in \cite{dmw} where the 
partition function of M-theory on $X_{10}\times \IS^1$ with fluxes 
classified by cohomology classes reproduced the partition function of type 
IIA string theory on $X_{10}$ with fluxes classified by K-theory. 
Our result is admittedly more modest since it has been derived just in 
a particular example; it would clearly be nice to develop a more general 
description. However, it also incorporates additional subtleties since the 
matching involves both bulk and localized fluxes. In this sense our 
matching suggests a generalization of results in the line of \cite{dmw} to 
situations with degenerate fibers in the M-theory geometry.

We have moreover extended our analysis to the situation with non-trivial 
type IIB 3-form fluxes, and shown there is a non-trivial interplay between 
quantization of bulk 3-form fluxes and world-volume magnetic fluxes. This 
manifests the rich interplay between fluxes and branes, and shows that it 
is in general non-trivial to combine D-brane configurations with bulk 
flux backgrounds. Extreme care must thus be applied to exploit the modular 
structure of the theory in model building. 

We would like to conclude with some additional comments:

\begin{itemize}

\item In the type IIB language, the K-theory constraints on the quantization 
conditions can be ultimately understood in terms of the global structure 
of the gauge group, namely in terms of the full set of gauge representations 
present in the spectrum of the theory. In this sense, the quantization 
conditions are more manifest in the F/M-theory picture, where states in a 
sufficient number of different representations are all realized on an 
equal footing, in terms of M2-branes wrapped on holomorphic 2-cycles. A 
similar comment applies to the heterotic dual of our model, where 
all these states are manifest in the perturbative spectrum as vectors in 
the Narain lattice. Hence, in heterotic theory, a straightforward analysis 
shows that there are no extra conditions beyond the quantization 
conditions following from the perturbative spectrum of the theory.

\item Our analysis has shown an equivalence between cancellation of 
(discrete K-theory) RR charges and proper quantization of $p$-form field 
strength fluxes. It is interesting that these two kinds of consistency 
conditions, which are often thought of as independent, are ultimately 
related. In retrospect, such relation is not unnatural, since both kinds 
of conditions are related to cancellation of global anomalies. 

\item 
Quantization conditions can be regarded as arising from the
requirement of absence of global anomalies (namely in the definition of 
the path integral of a suitably charged state).
An amusing feature of the extra K-theory condition (\ref{cancel1}) is 
that its violation leads to a very familiar global anomaly, namely 
a Witten $SU(2)$ global anomaly, on the world-volume of a suitable probe, 
a D7-brane wrapped on $\IT^4$. In the F/M-theory picture, this implies 
that, in this particular case, incorrectly quantized 4-form fluxes can be 
detected as Witten anomalies on suitable probes. It would be interesting 
to find other setups where flux quantization is related to such a 
familiar global anomaly.

\end{itemize}

Our main tool in deriving our results has been the extension of known 
string dualities to situations with non-trivial field strength fluxes. We 
expect many new interesting and surprising relations from such extensions.

\centerline{\bf Acknowledgements}

We thank F. Marchesano for useful conversations. 
A.M.U. thanks M.~Gonz\'alez for kind encouragement and  support. 
I.G.E thanks the CERN theory unit for hospitality during completion of 
this work. 
This work has been partially supported by CICYT (Spain), under project 
FPA-2003-02877, and the RTN networks MRTN-CT-2004-503369 `The Quest For 
Unification: Theory Confronts Experiment', and MRTN-CT-2004-005104 
`Constituents, Fundamental Forces and Symmetries of the Universe'.
The work by I.G.E. is supported by the Gobierno Vasco PhD fellowship 
program.


\end{document}